\documentclass[a4paper,10pt]{article}
\usepackage{amsfonts}
\usepackage{amssymb}  
\usepackage{latexsym}
\usepackage{enumerate}
\usepackage{amsmath}
\usepackage{amsthm}
\usepackage{graphicx}
\usepackage{multirow}
\usepackage{enumitem}
\usepackage{soul}
\usepackage{commath}
\usepackage{url}
\usepackage{hyperref}
\usepackage{makecell}
\usepackage{enumitem}
\usepackage{placeins}
\usepackage{pdflscape}
\usepackage{adjustbox}
\usepackage{latexsym}
\usepackage{parallel}
\usepackage{subfigure}
\usepackage[normalem]{ulem} 
\usepackage[flushright,sc]{caption2}
\newtheorem{example}{Example}

\begin{document}

\title{Rational expectations as a tool for predicting failure of weighted $k$-out-of-$n$ reliability systems}
\author{Jorgen-Vitting Andersen \\
{\small CNRS, Centre d'Economie de la Sorbonne, Universit\'e Paris 1 Pantheon-Sorbonne }
\\{\small Maison des Sciences Economiques,106-112 Boulevard de l\/'H\^opital, 75647, Paris Cedex 13 , France.}
\\
{\small E-mail: Jorgen-Vitting.Andersen@univ-paris1.fr}\\
\\
Roy Cerqueti \\
{\small Sapienza University of Rome -- Department of Social and Economic Sciences}\\
{\small Piazzale Aldo Moro, 5 -- 00185, Rome, Italy}\\
{\small Tel.: +39 06 4991 0511}\\
{\small E-mail: roy.cerqueti@uniroma1.it}\\
{\small and}\\
{\small London South Bank University -- School of Business}\\
{\small London SE1 0AA, UK}\\
\\
Jessica Riccioni\thanks{Corresponding author.} \\
{\small Sapienza University of Rome -- Department of Social and Economic Sciences}\\
{\small Piazzale Aldo Moro, 5 -- 00185, Rome, Italy}\\
{\small Tel.: +39 06 4991 0511}\\
{\small E-mail: jessica.riccioni@uniroma1.it} \\
} 
\maketitle
\begin{abstract}
Here we introduce the idea of using rational expectations, a core concept in economics and finance, as a tool to predict the optimal failure time for a wide class of weighted $k$-out-of-$n$ reliability systems. We illustrate the concept by applying it to systems which have components with heterogeneous failure times. Depending on the heterogeneous distributions of component failure, we find different measures to be optimal for predicting the failure time of the total system. We give examples of how, as a given system deteriorates over time, one can issue different optimal predictions of system failure by choosing among a set of time-dependent measures.

\end{abstract} \textbf{Keywords:} Rational expectations, weighted $k$-out-of-$n$ reliability systems, failure prediction, statistical measures.
\section{Introduction}


In reliability theory, it is important to be able to assess the failure times of systems with interconnected components and this leads to interesting questions in mathematical statistics and probability modelling. 
Here, we propose a stochastic model for evaluating and predicting the expected time of failure of a class of weighted $k$-out-of-$n$ systems. In particular, a system fails when a given number of its components fail, and the weight of a failed component is assumed to be opportunely reallocated to the surviving components -- under the so-called \emph{reallocation rule}.



To this end, we apply a Bayesian approach from a rational expectations perspective. 
In finance rational expectations are used to estimate the fundamental price of a given asset. The fundamental price is obtained as an expectation value of the price at time $t$ conditioned on all available information relevant for the given asset. Since people are assumed to be rational (i.e. to make trading decisions without biases) taking the expectation value should give the fair, or fundamental, price.  The idea then is that every time new information (relevant for the price of an asset) arrives (e.g. new information about earnings, interest rates, mergers, etc) this will impact the fundamental price of an asset. In similar fashion our idea is that every time new information arrives concerning individual component failure, this should lead to a new (optimal) prediction for the failure of the ensemble of components, i.e. system failure. We would like to point out that in finance ``all relevant information'' is in principle infinite and very vague to quantify: e.g. how does new information of the sickness of the CEO impact future earnings of a company? On the contrary, in our case ``information'' is crystal clear to quantify via the new values of component weights of the system. 
Specifically, we estimate the expected failure time of the given systems conditioned by the distribution of their weights, on the basis of a large collection of preliminary experiments in which such weights are observed. 





Generally, in the literature (see, for example, \cite{Sanyal}, \cite{Krishnamurthy}, \cite{Gokhale}, and \cite{Yacoub}), the failure of reliability systems is forecast by using scenario analysis based on the observation of real systems. Once the failure of a system has been observed, scenario analysis is performed to understand what happened to that specific system.

To predict the failure time of the investigated systems, referred to here as ``in-vivo" systems, we use all the information stored in a previously constructed ``information set". In particular, we create a catalogue of synthetic systems and observe their time evolution, from the beginning to failure. For each system in the catalogue, we store the time-varying weights -- i.e., the \emph{configurations} of the weights -- and the associated failure times. From a rational expectations perspective, we consider the arithmetic mean -- which plays here the role of the expected value -- of the failure times of the systems in the catalogue, under the constraint of a specific configuration of the weights. Then, 
we compare these expected values with the failure times of in-vivo systems with a similar configuration of the weights.
In doing so, we explore the informative content of the weights of the system components so that we can then use an extensive simulation approach to forecast failure times.

The method is inspired by the one proposed by \cite{Andersen, Sornette} for the prediction of failure time of the overall system, conditioned on the information revealed by the damage occurring up until the time at which the system is being evaluated -- i.e., according to the particular weight configuration.
This idea was in turn influenced by the method known as ``reverse tracing of precursors" (RTP) (see \cite{Keilis}, \cite{Shebalin}) for earthquake prediction based on seismicity patterns. 


There are two aspects to this. First, we discuss the initial distribution of the weights to identify which of them leads to more effective rational expectations-based predictions. Second, we hypothesize that the similarity of the weight configurations is captured by the similarity of one of their statistical indicators; accordingly, we test several statistical indicators to find out which has the highest predictive power. 

For the initial distribution of the weights, we follow the insights of \cite{Li}, \cite{Sarhan}, \cite{Asadi}, \cite{Eryilmaz}, \cite{Van}, and \cite{Zhang1} and compare five different initial distributions of the weights, viz., the uniform distribution in the unit interval and four types of Beta distribution, whose parameters cover the cases of symmetry and asymmetry to the left and to the right. 

We also exploit the existing literature to some extent for the analysis of the statistical indicators, considering the variance, skewness, kurtosis, Gini coefficient, and Shannon entropy. These statistical dimensions were chosen for their crucial informative content.

Concerning the use of moments in forecasting models, some authors consider the lowest moments of the distributions (the second in our case, i.e., the variance) to be more efficient than the higher moments (the third, skewness, and the fourth, kurtosis), which are attested to be less stable and reliable (see, e.g., \cite{Reijns}, \cite{Amari}, \cite{Ramberg}, and \cite{Kinateder}).

The usefulness of the Gini coefficient when dealing with failure prediction models is discusse by \cite{Ooghe}.

To the best of our knowledge, there are no contributions in the literature using the Shannon entropy for forecasting. Our analysis fills this gap. Importantly, the present paper provides a bridge between reliability theory and rational expectations.

The main message of this paper is the importance of rational expectations in the predictive context. As time goes by, the increase in the information available to predict failure times clearly improves the possibilities for prediction. So, after the initial phase of the trends, where the information available is highly random and there is no gain in rational expectations, the statistical indicators included in the analysis prove to be good predictive tools. The increase in knowledge we have about the system allows us to achieve better performance than the benchmark. In particular, at every time step, we observe an improvement in the error curves.

Our results agree with the existing literature regarding the initial weight distributions, confirming the predictive superiority of the negative-exponential distribution. Surprisingly, among the statistical measures, the Gini coefficient and the Shannon entropy give us the best predictive results, even though they are almost never used in forecasting models.

The rest of the paper is organized as follows.
Section \ref{sec2} contains a review of the main literature relating to our research. Section \ref{sec3} discusses the relevant reliability systems, with particular reference to their structure, the main properties of their components, and the failure rule. Section \ref{sec4} is devoted to the extensive simulations validating the theoretical proposal. Section \ref{sec5} contains the results and a critical discussion. The last section concludes and suggests future lines of research.

\section{Literature Review}
\label{sec2}

Here we review the basic literature that has inspired this research. There are two main themes: $k$-out-of-$n$ reliability systems and rational expectations.

The $k$-out-of-$n$ systems involve a variety of special cases and generalizations. However, we can distinguish two types of systems depending on the heterogeneity of their components. Specifically, systems with homogeneous components are ones where all the components contribute equally to the reliability of the system; otherwise a system is said to have heterogeneous components.

Scientific research on weighted $k$-out-of-$n$ systems was introduced in \cite{wuchen}. Studies of systems with homogeneous components and related applications can be found in \cite{Ge}, \cite{Boland}, and \cite{Milczek}. The heterogeneous case is more challenging. Cases with random and mutually independent weights can be found in \cite{CerquetiANOR}, \cite{Xie}, \cite{Li2}, \cite{Eryilmaz2},  \cite{Eryilmaz3}, \cite{Eryilmaz4}, \cite{Taghipour}, \cite{Zhang2}, \cite{Zhang3}, and \cite{Sheu}. In all the cases, prediction of the failure times of $k$-out-of-$n$ systems has seen the development of several methods and is widely debated among scholars.
It is worth mentioning some of the most relevant contributions:
\begin{itemize} 
\item Da Costa et al. \cite{daCosta} applied active redundancy or minimal standby redundancy depending on the nature of the treated systems, using a martingale approach. 
\item Eryilmaz \cite{Eryilmaz1} explored the mean residual lifetime as a fundamental characteristic to be used for dynamic reliability analysis. 
\item Wang et al. \cite{Wang} considered the reliability estimation of weighted $k$-out-of-$n$ multi-state systems. 
\item Zhang et al. \cite{Zhang} used a Monte Carlo simulation approach to confirm the accuracy of a model which assesses the reliability of a given system on the basis of the available information.
\end{itemize}


More generally, one can distinguish two main approaches to failure prediction in $k$-out-of-$n$ systems in the existing literature: a probabilistic approach that seeks to calculate the probability distribution of a system's failure time using techniques of stochastic calculus, and a Bayesian computational approach that estimates the average failure time of a system, conditioned on the description of a scenario in which the evolution of the given reliability system is observed.



In the former group, Oe et al. \cite{Oe} used autoregressive models to predict the failure of a stochastic system. For this purpose, they considered four types of performance index: quadratic distance of autoregressive parameter differences, variance of the residuals, Kullback information, and the distance of the Kullback information (divergence measure). Furthermore, Azaron et al. \cite{Azaron} used the reliability function for systems with standby redundancy. The system fails when all connections between input and output that are connected to the main components are broken. From a different viewpoint, Parsa et al. \cite{Parsa} introduced a new stochastic order based on the Gini-type index, showing how it could be used to gain information about the ageing properties of reliability systems, and thus establishing the characteristics of active or already failed components. 

In the probabilistic approach, a key role is played by the so-called \emph{coherent systems}, i.e., systems without irrelevant components; moreover, such systems certainly work when all the components are active and fail when all the components have failed. An important contribution here is due to Navarro and coauthors, who give some insight into the case of dependent components (see \cite{Navarro1}, \cite{Navarro2}, and \cite{Navarro3}).
In the same context, Gupta et al. \cite{Gupta} compared the residual lifetime and the inactivity time of failed components of coherent systems with the lifetime of a system that had the same structure and the same dependence. In doing so, the paper cited is particularly close to our own approach, in that it proposes a comparison between a test system and the investigated one -- as we do here, with the comparison between the investigated systems and those in the catalogue.
In contrast, Zarezadeh et al. \cite{Zarezadeh} investigated the joint reliability of two coherent systems with shared components, obtaining a pseudo-mixture representation for the joint distribution of the failure time.

This Bayesian approach includes many papers that estimate the average time to failure of these systems using asymmetric loss functions. 
Among them, Mastran \cite{Mastran} presented a procedure for exploring the connection between component failure and system collapse in the case of independent components.
Mastran and Singpurwalla \cite{Mastran2} also proposed examples of series systems with independent components and parallel systems with component interdependence by using prior data to build in this interdependence. Barlow and Heidtmann \cite{Barlow} modeled a combination of information between components and systems by exploiting lifetime data. Martz et al. \cite{Martz} proposed a very detailed procedure and explanatory examples for either test or prior data at three or more configuration levels in the system. Martz and Waller \cite {Martz2} extended \cite{Martz} for the particular cases of series and parallel subsystems. Regarding applications, we should mention \cite{Noortwijk}, who developed a Bayesian failure model for the observable deterioration characteristics in a hydraulic field. Gunawan and Papalambros \cite{Gunawan} presented a Bayesian-type reliability system model in the field of engineering. Kim et al. \cite{kim} explored deteriorating systems by conditioning on monitoring data based on three-state continuous time homogeneous Markov processes. Along the same lines, \cite{aktekin} presented Markov chain Monte Carlo methods in the area of software reliability to investigate the failure rate of systems with components that change stochastically.
Regarding Bayesian methods, it is also worth mentioning \cite{Bhattacharya}, \cite{El-Sayyad},  \cite{Canfield}, \cite{Varian}, \cite{Zellner}, and \cite{Basu}. 



In line with several of the above-mentioned studies, the weighted $k$-out-of-$n$ systems presented here have components with an inner dependence structure. The failure of a given system is assumed to depend on the number and importance of the components. The approach we follow is Bayesian. Indeed, our aim is to estimate the failure time of a system by using the information collected in an observed catalogue as the prior. 
In doing so, 
we introduce a new element into reliability theory by adopting a rational expectations perspective.

The way expectations are created is a classical theme in the economic debate. Indeed, in the modern theory of behavioral economics, agents are divided into two different categories depending on the hypothesis they use to form economic expectations: those who follow the hypothesis of adaptive expectations, and others who instead follow the hypothesis of rational expectations. Under the first hypothesis, future economic decisions are made according to what happened in the past (see, e.g., \cite{Friedman} and \cite{Chow}). In the context of rational expectations, however, future outcomes are 
computed as the conditional expectation of the observed realizations of the quantity to be predicted given the available information (see the breakthrough contributions by \cite{Muth}, \cite{Lucas}, \cite{Sargent}, \cite{Sargent1}, and \cite{Barro}). 

%
%

The rational expectations hypothesis is a fundamental assumption in many theoretical models, with implications for economic analysis, and thanks to the increasing accessibility of big data in recent years, studies have been carried out on the use of rational expectations to identify prediction errors in large samples. So, rational expectations are important in any situation in which the agents' behavior is influenced by expectations (see \cite{Maddock}).


In the context of forecasting, we mention \cite{atici} who applied Cagan's model of hyperinflation on discrete time domains. From a different perspective, Becker et al. \cite{becker} compared the rational expectations hypothesis with the bounds and likelihood heuristic to explain average forecasting behavior.

The rational expectations perspective proposed here is used to forecast the failure time of a reliability system, given the available information. Specifically, we check whether the conditional expectation of the realizations in the catalogue of such a quantity given a peculiar state of the weights leads to a suitable identification of the failure time of a system with the same weight configuration.


\section{The reliability system}
\label{sec3}

We consider a probability space $(\Omega,\mathcal{F},\mathbb{P})$
containing all the random quantities used throughout the paper. We denote the expected value operator related to the probability measure $\mathbb{P}$ by $\mathbb{E}$.
\newline
We denote the \emph{reliability system} -- or, simply, the \emph{system} -- by $\mathbf{S}$, and assume that it comprises $n$ \emph{components} denoted by $C_{1},\dots,C_{n}$ and
collected in a set $\mathcal{C}$.
\newline
As we will see, in our framework the system can be considered to be of weighted $k$-out-of-$n$ type in the sense that it fails when some of its components fail. 
\newline
The \emph{state} of $\mathbf{S}$ is a binary quantity. If the system
is \emph{active} and works, then its state is 1. Otherwise, the
state of $\mathbf{S}$ is 0, and the system is said to have
\emph{failed}. The state of $\mathbf{S}$ evolves in time and is
denoted by $Y(t)$ at time $t \geq 0$. At the
beginning of the analysis (time $t=0$), the system is naturally
assumed to be in state 1.
\newline
Analogously, the state of the $j$-th component $C_j$ at time $t$ is
denoted by $Y_j(t)$, and it takes value 1 when $C_j$ is active and 0
when $C_j$ has failed. At time $t=0$ we have $Y_j(0)=1$, for each
$j=1, \dots, n$.

\subsection{The structure of the system}

To express the dependence of the state of $\mathbf{S}$ on the states
of its components, we simply introduce a function $\phi:\{0,1\}^n
\rightarrow \{0,1\}$
\begin{equation}
\label{phi} Y(t)=\phi(Y_1(t), \dots, Y_n(t)). \end{equation}

In reliability theory, $\phi$ is usually called the
\emph{structure function} of the system.

The elements of $\{0,1\}^n$ are called configurations of the
states of the components of the system or, briefly,
\emph{configurations}.

The function $\phi$ in \eqref{phi} has the role of clustering the set of
configurations into two subsets: those leading to failure (F)
of the system and those associated with the not-failed (NF) system.
Thus, we say that $K_F\subseteq \{0,1\}^n$ is the collection of
configurations such that $\phi(x_F)=0$, for each $x_F \in K_F$, while
$K_{NF}\subseteq \{0,1\}^n$ is the collection of configurations such
that $\phi(x_{NF})=1$, for each $x_{NF} \in K_{NF}$. By definition,
$\{K_F,K_{NF}\}$ is a partition of $\{0,1\}^n$.

We may reasonably assume that the system is coherent and that the following three conditions are satisfied: 

First, $(0, \dots, 0)\in K_F$ and $(1, \dots, 1)\in K_{NF}$. This
condition means that, when all the components of the system are
active (not active), then the system as a whole is also active (not active).

Second, $\phi$ is non-decreasing with respect to its components.
This has an intuitive explanation: the failure of one of the
components of the system might worsen the state of the system and
cannot improve it.

Third, each component is able to determine the failure of the
system. Formally, this condition states that, for each $j=1,\dots, n$,
there exists $\left(y_1, \dots, y_{j-1},y_{j+1}, \dots, y_{n}\right)
\in \{0,1\}^{n-1}$ such that $(y_1, \dots, y_{j-1},1,y_{j+1}, \dots,
y_{n}) \in K_{NF}$ and $(y_1, \dots, y_{j-1},0,y_{j+1}, \dots,
y_{n}) \in K_{F}$.

\subsection{Components and Weights}

We now note three natural assumptions about the components of the system, inspired by standard reliability theory: first, the various components of the system don't all have the same relevance; second, the
components of the system are interconnected and exhibit different
levels of interconnection; third, relevance and interconnection
levels change over time, with the changing status of
the components of the system. We now spell this out.

For each $j=1, \dots, n$ and $t \geq 0$, the \emph{relative
	importance of the component} $C_j$ over the entire system at time
$t$ is measured by $\alpha_j(t)$, where $\alpha_j:[0, +\infty)
\to [0,1]$ and $\sum_{j=1}^n \alpha_j(t)=1$, for each $t$.

For each $t \geq 0$, we collect the $\alpha(t)$'s in a time-varying
vector $\mathbf{a}(t)=(\alpha_j(t))_{j}$, where
\begin{equation}
\label{a} \mathbf{a}:[0,+\infty) \to [0,1]^n\qquad \rm{such
	\,\,that}\qquad t \mapsto \mathbf{a}(t).
\end{equation}

If a component is not active at time $t$, then its relevance for the
system is null. Moreover, each active component has positive
relative relevance, i.e., the system does not contain irrelevant
active components. Formally,
\begin{equation}
\label{alphaY} \alpha_j(t)=0 \Leftrightarrow Y_j(t)=0.
\end{equation}
Condition \eqref{alphaY} is useful, in that it allows us to describe
the status of the system's components directly through the
$\alpha$'s.

For each $j=1, \dots, n$, the relative relevance of $C_j$ varies with the variation of the state of each of the system's components. Once a component fails, it disappears from the
reliability system -- i.e., its relative relevance becomes
null -- and the relative relevances of the components of the
remaining active ones are modified on the basis of a suitably
defined \emph{reallocation rule}.



\begin{figure} [h]
	
	\centering
	
	\includegraphics[width=10cm]{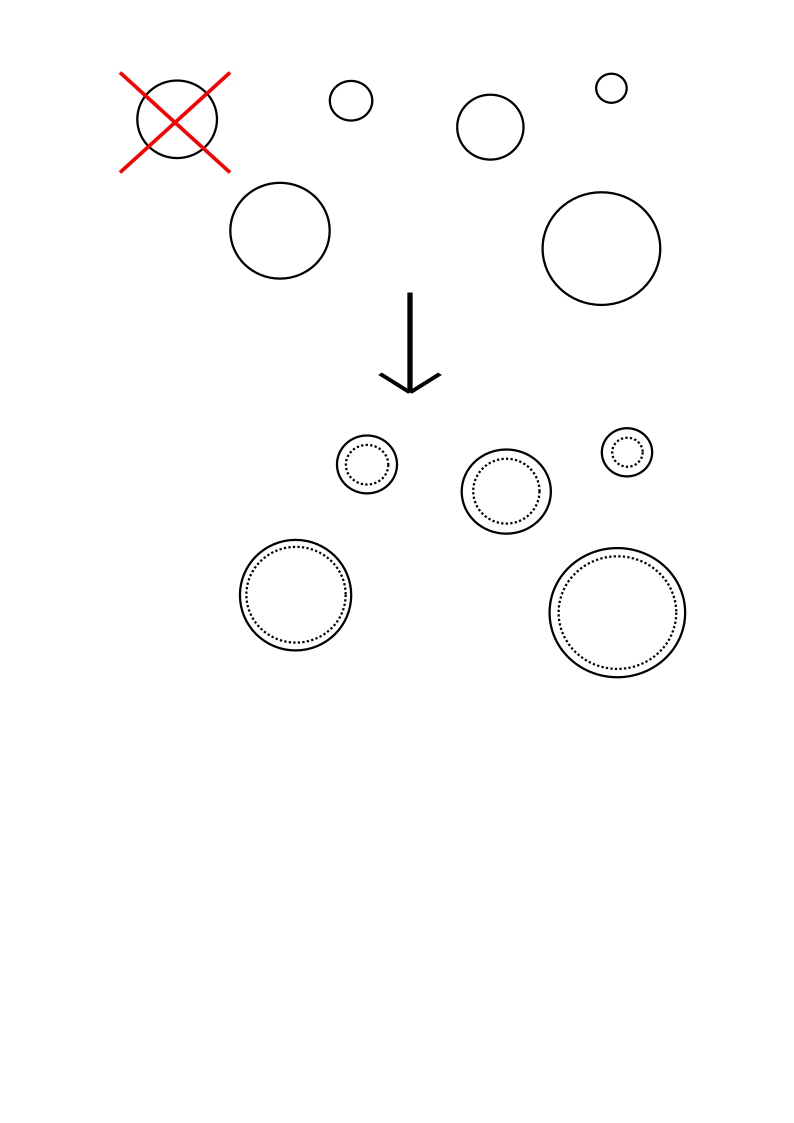}
	
	\captionstyle{normal}
	\caption{Blob graph representing the proportional reallocation rule described in Example \ref{proportional}: [Top] A component has failed and is deleted from the system. Then, [Bottom] The relevance of the failed component is reallocated over the remaining active components in proportion to their $\alpha$ values? (bubble size) before the failure.}
	\label{blobgraph}
\end{figure}

\newpage

The next example proposes a way to build a reallocation rule.

\begin{example}
	\label{proportional}
	Consider a system $\mathbf{S}$ whose component set is
	$\mathcal{C}=\{C_1,C_2,C_3,C_4,C_5\}$.
	\newline
	Assume that, at time $t=0$, we have $\alpha_1(0)=0.1$,
	$\alpha_2(0)=0.15$, $\alpha_3(0)=0.3$, $\alpha_4(0)=0.2$,
	$\alpha_5(0)=0.25$.
	
	Now, suppose that the first failure of one of the components of the
	system occurs at time $t=7$, when $C_3$ fails. Of course,
	$\alpha_j(t)=\alpha_j(0)$, for each $t \in [0,7)$ and $j=1,2,3,4,5$.
	Moreover, $\alpha_3(7)=0$.
	
	We consider a specific reallocation rule which states that the
	relevance is reallocated over the remaining active components
	in proportion to their $\alpha$ values before the failure (see Figure \ref{blobgraph}). This means
	that
	$$\alpha_1(7)=\frac{0.1}{0.1+0.15+0.2+0.25},\,\,\alpha_2(7)=\frac{0.15}{0.1+0.15+0.2+0.25},$$
	$$\alpha_3(7)=0,
	\,\,\alpha_4(7)=\frac{0.2}{0.1+0.15+0.2+0.25},\,\,\alpha_5(7)=\frac{0.25}{0.1+0.15+0.2+0.25}.$$
	
	In general, if $\tau_1, \tau_2$ are the dates of two consecutive
	failures, with $\tau_1<\tau_2$, we have
	$$
	\alpha_j(\tau_2)=\frac{\alpha_j(\tau_1)\mathbf{1}_{\{Y_j(\tau_2)=1\}}}{\sum_{i=1}^5\alpha_i(\tau_1)\mathbf{1}_{\{Y_i(\tau_2)=1\}}},
	\qquad j=1,2,3,4,5. $$
	
	The $\alpha$'s are step functions, with jumps each time one of the components fails.

\end{example}

Regarding the \emph{interconnections among the components},
we define their time-varying relative levels using functions $w_{ij}:[0, +\infty) \to [0,1]$, for each $i,j=1, \dots, n$, so
that $w_{ij}(t)$ is the relative level of the interconnection
between $C_i$ and $C_j$ at time $t \geq 0$. We assume that the arcs in the resulting network 
are
oriented, so that in general $w_{ij}(t) \neq w_{ji}(t)$, for each
$t$. Moreover, by construction, $\sum_{i,j=1}^n w_{ij}(t)=1$, for
each $t$. We also assume that self-connections do not exist in our
network, i.e., $w_{ii}(t)=0$, for each $i$ and $t$.

For each $t \geq 0$, the $w(t)$'s are collected in a time-varying
matrix $\mathbf{w}(t)=(w_{ij}(t))_{i,j}$, with
\begin{equation}
\label{w} \mathbf{w}:[0,+\infty) \to [0,1]^{n\times n}\qquad
\rm{such \,\,that}\qquad t \mapsto \mathbf{w}(t).
\end{equation}

If $C_i$ is a non-active component at time $t$, then
$w_{ij}(t)=w_{ji}(t)=0$, for each $j=1,\dots, n$. This condition
simply formulates the idea that a failed component is disconnected from the
system. It suggests that the failure of a component
might generate disconnections among the components of the system.

The behavior of the $w$'s is analogous to that of the $\alpha$'s.
In this case, too, the relative levels of interconnections change
when one of the components of $\mathbf{S}$ changes its state, and
there is a reallocation rule for the remaining levels of
interconnection.

We denote the whole set of reallocation rules for the weights on nodes and
arcs by $\mathcal{R}$. 
Therefore, a natural rewriting of the system $\mathbf{S}$ with
components in $\mathcal{C}$ and reallocation rule $\mathcal{R}$ at time $t$ is
\begin{equation}
\label{S} \mathbf{S}(t)=\{\mathbf{a}(t), \mathbf{w}(t)\}.
\end{equation}
Notice that (\ref{S}) highlights the observable features of the
system with a given set of components and a specific reallocation
rule, i.e., the weights on the nodes and on the arcs. Thus, according
to (\ref{S}), we can refer to $\{\bar{\mathbf{a}},
\bar{\mathbf{w}}\}$ as an \emph{observation} of the system at a given time, where
$\bar{\mathbf{a}} \in [0,1]^n$ and $\bar{\mathbf{w}} \in [0,1]^{n
	\times n}$.

When needed, we will conveniently remove the dependence on $t$ from the quantities in (\ref{S}).

\subsection{Failure of the system}

As mentioned, time $t=0$ represents \emph{today} -- the 
point at which we begin to observe the evolution of the system. Since the system is coherent in the sense that there are no irrelevant components, at time
$t=0$ all the components are active and the system works.
The failure of the system is then a random event, which occurs when
the system achieves one of the configurations belonging to $K_F$.

We define the \emph{system lifetime} by
\begin{equation}
\label{mathcalT} \mathcal{T}:=\inf\{t\geq 0|\phi(Y_1(t), \dots,
Y_n(t))=0\}.
\end{equation}
Analogously, the $n$-dimensional vector of \textit{component
	lifetimes} is $\mathbf{X}=(X_{1},\dots,X_{n})$, where
\begin{equation}
\label{Xj} X_{j}=\inf\{t>0\,|\,Y_{j}(t)=0\}.
\end{equation}
To be as general as possible, we assume that the failure lifetimes of the components of the system $\{X_{1}, \dots, X_{n}\}$ are not independent random variables and do not share the same distribution. In fact, for each component failure, the $\alpha$'s and $w$'s change in accordance with the reallocation rule $\mathcal{R}$; this also modifies the probability of subsequent failures of the system components in the very natural case where failures depend on the weights. 
\newline
Moreover, we can reasonably assume that the failure of the system coincides
with the failure of one its components. 
%
%


To fix ideas, we provide an example.

\begin{example}
	Assume that $\mathcal{C}=\{C_1, C_2,C_3,C_4,C_5\}$ and
	$$
	\mathbf{a}(0)=(0.1,0.5,0.2,0.1,0.1), \qquad \mathbf{w}(0)=\left(
	\begin{array}{ccccc}
	0 & 0 & 0.1 & 0.1 & 0 \\
	0 & 0 & 0.1 & 0 & 0 \\
	0.1 & 0.1 & 0 & 0.1 & 0.05 \\
	0.1 & 0 & 0.1 & 0 & 0.05 \\
	0 & 0 & 0.05 & 0.05 & 0 \\
	\end{array}
	\right)
	$$
	
	Suppose that the reallocation rules $\mathcal{R}$ for relative
	relevance and interconnection levels are of proportional type, as in
	Example \ref{proportional}. Such reallocations are implemented if the
	system has not failed.
	
	Furthermore, assume that the failure of a component has a twofold nature: on the one hand, it can be brought about by an idiosyncratic shock; on the other, it can be driven by the failure of the other components. Specifically, we hypothesize that, if a given component fails, then other
	components that are only connected to that component will fail as well, independently of
	their levels of interconnection. In contrast, the idiosyncratic shocks are assumed to be captured by a Poisson
	process with parameter $\lambda$ -- giving the timing of the
	failures -- jointly with a uniform process over $\mathcal{C}$,
	independent of the Poisson process, which identifies the failed
	component.
	
	Moreover, suppose that the system fails at the first time in which
	components with aggregated relative relevance greater than 0.4 fail.

	Now, suppose that the first failure is observed at time $t=8$, when
	$C_2$ fails. Then, automatically, $C_3$ fails as well, since it is
	connected only to $C_2$. The aggregate relative relevance before the
	failures is $\alpha_2(8^-)+\alpha_3(8^-)=0.5+0.2>0.4$, and the
	system fails.

\end{example}


To compute the expected failure time of the system under a rational expectations approach, we use the information contained in the specific values of the weights at time $t$, namely $(\mathbf{a}(t), \mathbf{w}(t))$. 
Specifically, we compute the expected value of $\mathcal{T}$ at time $t$, 
conditioned on the specific values of the weights $(\mathbf{a}(t), \mathbf{w}(t))$.

%

If $RE_t$ is the value of the rational expectations prediction, issued at time $t$, of the failure time
$\mathcal{T}$, given all the possible observations of the system, we have

\begin{equation}
\label{RE}
RE_t=\left\{\mathbb{E}\left[\mathcal{T}\,|\,(\bar{\mathbf{a}}(t), \bar{\mathbf{w}}(t))
\right]\,:\,(\bar{\mathbf{a}}(t), \bar{\mathbf{w}}(t))\in [0,1]^n \times [0,1]^{n \times n}
\right\}.
\end{equation}
The formula (\ref{RE}) gives the expected value of the lifetime of
$\mathbf{S}(t)$ conditioned on the specific observations of the state of the system at time $t$, viz., $(\bar{\mathbf{a}}(t), \bar{\mathbf{w}}(t))$. We note the link to rational expectations used in finance, with the left hand side of (\ref{RE}) corresponding to the fundamental price of an asset, which can be obtained as an expectation value of the price at time $t$ conditioned on all available information relevant for the given asset. in finance, the idea is that every time new information arrives (e.g. new information about earnings, interest rates, mergers, etc) this will impact the fundamental price of an asset. In similar fashion our idea is that every time new information arrives concerning individual component failure, this should lead to a new (optimal) prediction for the failure of the ensemble of components, i.e. system failure. We would like to point out that in finance ``all relevant information'' is in principle infinite and very vague to quantify: e.g. how does new information of the sickness of the CEO impact future earnings of a company? On the contrary, in our case ``information'' is crystal clear to quantify via the new values of $(\bar{\mathbf{a}}(t), \bar{\mathbf{w}}(t))$.

In order to calculate, for a given system at a given time, the conditioned expected value in (\ref{RE}) for the investigated systems, referred to here as ``in vivo" systems, the central trick is to first create a ``catalogue'' of information. 
To produce the catalogue, we generate a set of $M$ systems and we follow their lives from the beginning to failure. For each time $t$, we record the state of each system of the catalogue $(\bar{\mathbf{a}}(t), \bar{\mathbf{w}}(t))$ and the failure time  $\mathcal{T}$, i.e., we effectively record the pair ($(\bar{\mathbf{a}}(t), \bar{\mathbf{w}}(t))$, $\mathcal{T}$). In this way, we create an \emph{``information set"} $\mathbf{I}^M$. Now, each in vivo system at each time step $t$ presents a specific configuration $(\bar{\mathbf{a}}(t), \bar{\mathbf{w}}(t))$, and we can predict its failure time $RE_t$ using (\ref{RE}) and referring to the systems in $\mathbf{I}^M$.

In practice, the use of (\ref{RE}) to create a large enough information set on a computer requires a lot of CPU time and storage space. We reduce the dimensionality of the problem by introducing a function $f$ that maps the weights $(\bar{\mathbf{a}}(t), \bar{\mathbf{w}}(t))$ onto real numbers. 
We then replace (\ref{RE}) by
\begin{equation}
\label{RE1}
RE^f_t=\left\{\mathbb{E}\left[\mathcal{T}\,|\,f(\bar{\mathbf{a}}(t), \bar{\mathbf{w}}(t))
\right]\,:\,(\bar{\mathbf{a}}(t), \bar{\mathbf{w}}(t))\in [0,1]^n \times [0,1]^{n \times n}
\right\}.
\end{equation}
The idea now is to test our method using the variance, skewness, kurtosis, Gini coefficient, and Shannon entropy as the function $f$. The next section details the simulation procedure.


\section{Simulation experiments}
\label{sec4}
%
%
We now present the simulation procedure used to test our methodological proposal.

\subsection{Specifying the systems}

The procedure for the failure of components works in a stepwise form. 
The system is assumed to fail the first time the number of failed components exceeds $N/2$. 

The reallocation rule is of proportional type, as in Example \ref{proportional}, so the relevance of the failed component is reallocated over the remaining active components in proportion to their $\alpha$ values before the failure. Such a reallocation rule, together with the condition for failure described above, clearly indicates the stochastic interdependence of the lifetimes of the system components. This said, for the sake of simplicity, we keep this interdependence implicit by setting the links among the components equal, i.e., at any given time $t$, any active component $C_i$ is connected with the same strength $w_{ij}$ to any other active component $C_j$, whence all the entries in $\mathbf{w}(t)$ can be taken as equal to unity, for each time $t$. We can thus remove any further reference to the matrix $\mathbf{w}$.

We now consider the condition for the failure of a component. 
At a generic time $t$, one component, say $C_j$, is selected at random, i.e., from a uniform distribution on the set of active components, as the candidate failed component. Then, a random number $r$ is sampled from a uniform distribution $U(0,1)$. If $\alpha_j(t) > r$, the component $C_j$ fails at time $t$, and $\alpha_j(s)=0$, for each $s>t$. On the other hand, if $\alpha_j(t) \leq r$, the component $C_j$ does not fail. This procedure is then reiterated at time $t+1$ and so on, until the system fails.

\subsection{Stepwise description of the procedure}
%
%

We consider two different sets of systems, \emph{information set} systems and \emph{in-vivo} systems. The in-vivo systems are the ones we wish to predict, the information set systems are the systems we use to create an information set, from which we can issue a prediction of an in-vivo system via Eq. \ref{RE1}. We first create an information set  by letting a certain number of systems $M$ fail. For each time $t$ we record the state of the system $f(\bar{\mathbf{a}}(t))$, and when the system fails, at say time  $\mathcal{T}$, we record the pair ($f(\bar{\mathbf{a}}(t)$),$\mathcal{T}$). Repeating this procedure for $M$ different systems, we thereby create an \emph{``information set"} $\mathbf{I}(t)^M_f$. The idea behind the rational expectations approach is then, for each in vivo system at each time step $t$, to use the systems in $\mathbf{I}(t)^M_f$ with the same information to give an averaged evaluation of the prediction time via (\ref{RE1}).

We present the simulation procedure in a stepwise form:
\begin{enumerate}
	\item For each function $f$, we build the \emph{information set} $\mathbf{I}_f=(\mathbf{I}(t)^m_f\,:\, t \geq 0; \, m=1,\dots, M)$ by creating and following the lives of $M$ systems from time $t=0$ until they fail. For each time $t$ we record the state of the $m$-th system $f(\bar{\mathbf{a}}_m(t))$, and when the system fails, at say time  $\mathcal{T}_m$, we record the pair ($f(\bar{\mathbf{a}}_m(t)$),$\mathcal{T}_m$). We do this for each $m=1, \dots, M$.
	%

	\item We follow the same procedure by creating $X$ in-vivo systems and following their lives from the beginning at time $t=0$ until they fail. 
	For each time $t$, we record the state of the $x$-th system $f(\bar{\mathbf{a}}_x(t))$, and its failure time $\mathcal{T}_x$, i.e., the pair ($f(\bar{\mathbf{a}}_x(t)$),$\mathcal{T}_x$). 
	
		
%
%

	\item We now compute the rational expectations in (\ref{RE1}) on the basis of the information set $\mathbf{I}_f$. 
	
	\begin{itemize}
		\item First, we state and check a tolerance threshold condition. 
		
		Specifically, we fix a tolerance level $\Theta >0$; then, for each in vivo system $\bar{x}=1, \dots, X$, time $\bar{t} =0,1,\dots, \mathcal{T}_{\bar{x}}$, and observed configuration $f(\bar{\mathbf{a}}_{\bar{x}}(t)$, we identify the systems in the catalogue with label $m \in \{1, \dots, M\}$ such that the following \emph{Condition} holds:
		\begin{equation}
		\label{condition}
		|f(\bar{\mathbf{a}}_{\bar{x}}(\bar{t})- f(\bar{\mathbf{a}}_m(\bar{t})| < \Theta.
		\end{equation}

We denote the number of systems in the catalogue for which (\ref{condition}) is satisfied by $m[f(\bar{\mathbf{a}}_{\bar{x}}(\bar{t})]$. Hypothetically, one might have $m[f(\bar{\mathbf{a}}_{\bar{x}}(\bar{t})]=0$. In this unlucky case, the catalogue does not provide complete information about the configurations of the systems. To avoid such an inconsistency, we have reasonably selected values of $\Theta$ and $M$ large enough to guarantee that $m[f(\bar{\mathbf{a}}_{\bar{x}}(\bar{t})]>0$ for each $\bar{x}$ and $\bar{t}$ (see the next subsection, where we introduce the parameter set we used here).

\item To apply (\ref{RE1}), we compute the arithmetic mean of the failure times of the systems of the catalogue satisfying (\ref{condition}), so that

\begin{equation}
		\label{condition2}
		\mathbb{E}\left[\mathcal{T}\,|\, f(\bar{\mathbf{a}}_{\bar{x}}(\bar{t})\right]= \frac{1}{m[f(\bar{\mathbf{a}}_{\bar{x}}(\bar{t})]} \sum_{m=1}^M \mathcal{T}_m \cdot 1(m \text{ satisfies }(\ref{condition})), 
		\end{equation}
where $1(\bullet)$ the indicator function with value 1 if $\bullet$ is satisfied and 0 otherwise.

		
	\end{itemize} 
	
\item In order to assess the quality of such predictions, we proceed as follows:
\begin{itemize}
\item First, at each time $\bar{t}$, we consider the 10th, 50th, and 90th percentiles of the distributions of the observed configurations of the catalogue $\bar{\mathbf{a}}_m(\bar{t})$'s, with $m=1, \dots, M$. We denote these percentiles by $\bar{\mathbf{a}}^{(10)}(\bar{t})$, $\bar{\mathbf{a}}^{(50)}(\bar{t})$, and $\bar{\mathbf{a}}^{(90)}(\bar{t})$, respectively. 
\item We assign system ${\bar{x}}$ at time $\bar{t}$ with configuration $\bar{\mathbf{a}}_{\bar{x}}(\bar{t})$ to the percentile $\bar{\mathbf{a}}^{(per(\bar{x}))}(\bar{t})$ if and only if the following condition is satisfied: 
\begin{equation}
	\label{conditiondistance}
	|\bar{\mathbf{a}}_{\bar{x}}(\bar{t}) - \bar{\mathbf{a}}^{(per(\bar{x}))}(\bar{t})| <  0.8\times std(\bar{t}),
	\end{equation}
where $per=10,50,90$, $per(\bar{x})$ is the percentile related to $\bar{x}$, and $std(\bar{t})$ is the standard deviation of the configurations of the in vivo systems at time $\bar{t}$, namely the $\bar{\mathbf{a}}_{x}(\bar{t}) $'s with $x =1, \dots, X$. 
\item We compute
\begin{equation}
		\label{condition3}
		\tilde{\mathbb{E}}\left[\mathcal{T}\,|\, f(\bar{\mathbf{a}}_{\bar{x}}(\bar{t})\right]= \frac{1}{|per(\bar{x})|} \sum_{x=1}^X \mathcal{T}_x \cdot 1(x \text{ satisfies }(\ref{conditiondistance})),
		\end{equation}
where $|per(\bar{x})|$ is the cardinality of the set of the percentile $per(\bar{x})$.

\item We compute the error at time $\bar{t}$ and at a given percentile $per$ -- i.e., the difference in absolute value between the term in (\ref{condition2}) and the one in condition (\ref{condition3}) conditioned on the percentile $per$:
\begin{equation}
	\label{error}
	E_{RE}|per(\bar{t})=\left| \mathbb{E}\left[\mathcal{T}\,|\, f(\bar{\mathbf{a}}_{\bar{x}}(\bar{t})\right] -\tilde{\mathbb{E}}\left[\mathcal{T}\,|\, f(\bar{\mathbf{a}}_{\bar{x}}(\bar{t})\right] \right|.	\end{equation}
In this way, at each time $\bar{t}$ we obtain three different distributions of errors conditioned on the percentiles:
	\begin{itemize}
		\item $E_{RE}|10(\bar{t})$
		\item $E_{RE}|50(\bar{t})$
		\item $E_{RE}|90(\bar{t})$
	\end{itemize}

	\item We also compare the distributions in (\ref{error}) with a naive benchmark error $E_B$ given by the errors made without the use of rational expectations, as follows:

	\begin{equation}
	\label{benchmark}
	E_B=\frac{1}{X}\sum_{x=1}^X \left|\frac{1}{M} \sum_{m=1}^M \mathcal{T}_m - \mathcal{T}_x \right| 
	\end{equation}

As the formula (\ref{benchmark}) states, the performance of the naive benchmark prediction is obtained by constantly issuing the failure time of a given in vivo system, using the averaged failure time of the systems in the information set. The benchmark prediction is a natural measure for making forecasts without using rational expectations; indeed, it assumes that the failure prediction is given by the average of the failures of the systems in the catalogue.

\item For the purposes of comparison, we normalize all the times at the percentile level, so that for each percentile $per$, the maximum time over all the in vivo systems in which $\bar{\mathbf{a}}^{(per)}$ is observed is unity, while the minimum time is 0.
%
%
%
%
%
%
	
\end{itemize}

\end{enumerate}

\subsection{Setting the parameters}


The parameters for the catalogue and in vivo systems are: $n=10$, $M=5000$ and $X=5000$. This choice leads to satisfactory results without resulting in too much computational complexity.

\begin{figure}[h]
	\centering
	\includegraphics[scale = 0.6, trim = 0cm 0cm 0cm 0cm]{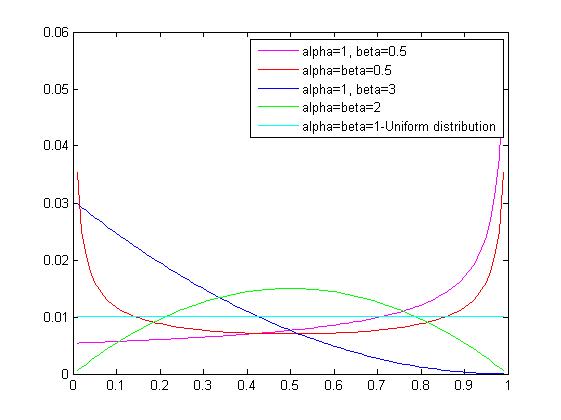}
	\captionstyle{normal}
	\caption{Beta distribution with four different combinations of shape parameters, together with the uniform distribution}
	\label{beta}
\end{figure}

The initial distribution of the weights in $\mathbf{a}(0)$ is assumed to be generated from different types of random variable with particular characteristics. Specifically, we take:

\begin{itemize}
	
	\item Uniform distribution $U(0,1)$.
	\item Some cases of the two-parameter Beta distribution $B(a,b)$. This distribution has support in (0,1); moreover, depending on the values assigned to the shape parameters $a$ and $b$, the behavior of the density function of $B(a,b)$ can be of different type. We consider four combinations of shape parameters (see Fig. \ref{beta}):
	\begin{itemize}
		\item $a=1$ and $b=3$, which is an asymmetric distribution more concentrated over the values close to zero;
		\item $a=b=0.5$, which is a symmetric distribution bimodal over the endpoints 0 and 1;
		\item $a=b=2$, which corresponds to a platykurtic symmetric distribution centered in 0.5;
		\item $a=1$ and $b=0.5$, which is an asymmetric distribution on the right with a high concentration of values close to 1.
	\end{itemize}
Note that a $B(a,b)$ with $a=b=1$ is a special case, corresponding to $U(0,1)$.
\end{itemize}

\newpage

The tolerance level $\Theta$ plays a key role in the analysis, as (\ref{condition}) suggests.
We evaluated three different tolerance levels, taking $\Theta=0.005, 0.05, 0.5$.

It turned out that the results were scarcely affected by the tolerance levels. Thus, we present here only the case $\Theta=0.05$.

\section{Results and discussion}
\label{sec5}

\begin{figure}
	\begin{minipage}[b]{3cm}
		\centering
		\includegraphics[width=5cm]{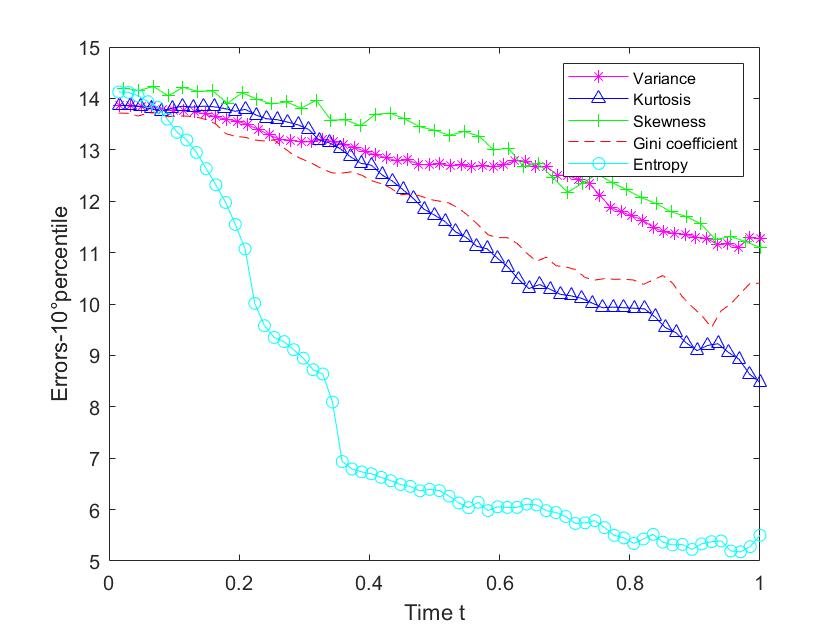}
			{\small A)\par}
	\end{minipage}
	\ \hspace{5mm} \hspace{10mm} \
	\begin{minipage}[b]{3cm}
		\centering
		
		\includegraphics[width=5cm]{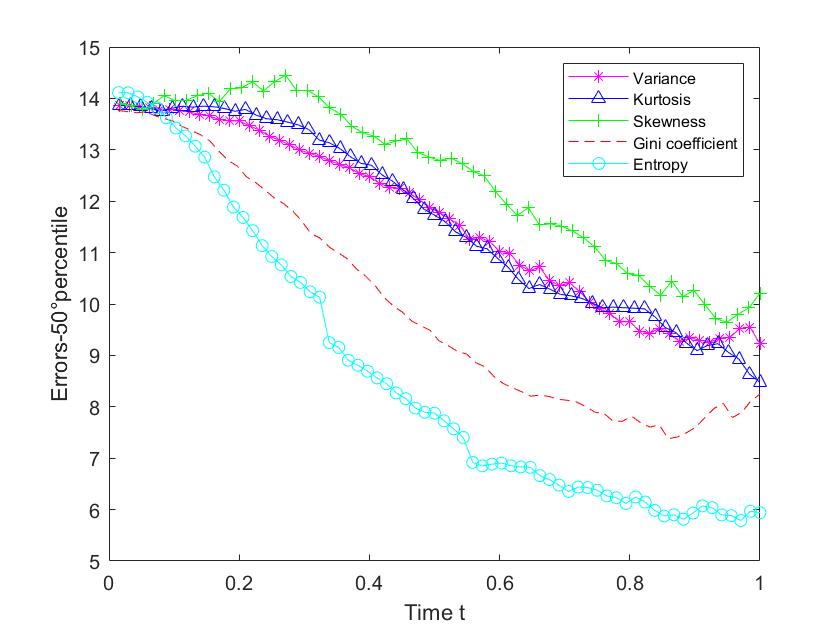}
		{\small B)\par}
	\end{minipage}
	\ \hspace{5mm} \hspace{10mm} \
	\begin{minipage}[b]{3cm}
		\centering
		\includegraphics[width=5cm]{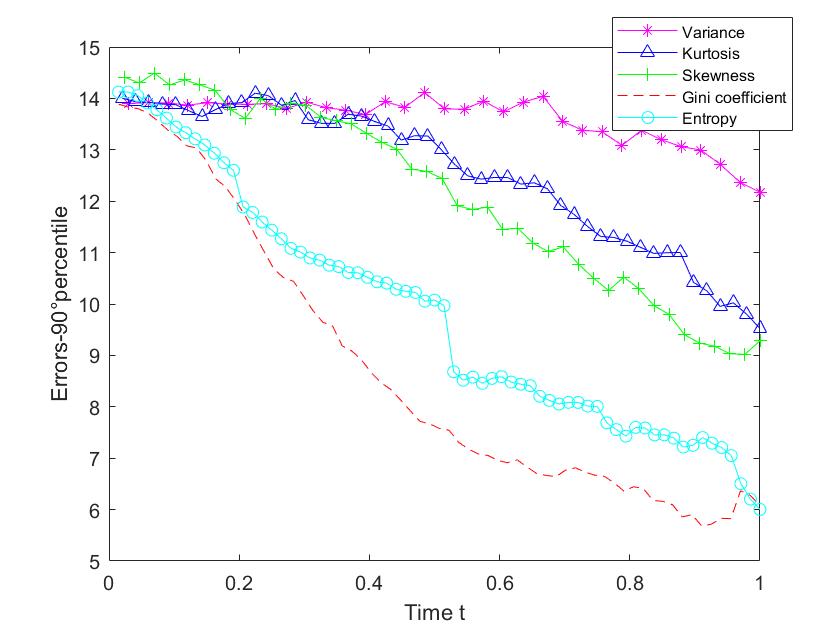}
		{\small C)\par}
	\end{minipage}
	
	\captionstyle{normal}
	\caption{Uniform distribution of initial weights. The $y$-axis shows the prediction error (absolute value) and the $x$-axis shows the time normalized so that failure occurs at $t=1$ in order to compare performance of the predictions across systems. Symbols correspond to the five different choices of function $f$ as indicated in the key. Panel A) shows the errors of the $10^{th}$ percentile of the distribution for each function calculated using the information set $\mathbf{I}^M$. Panels B) and C) show likewise the errors of the $50^{th}$ and $90^{th}$ percentiles. The horizontal line in each panel shows the performance of the benchmark prediction}
	\label{2}
\end{figure}

\begin{figure}
\begin{minipage}[b]{3cm}
\centering
\includegraphics[width=5cm]{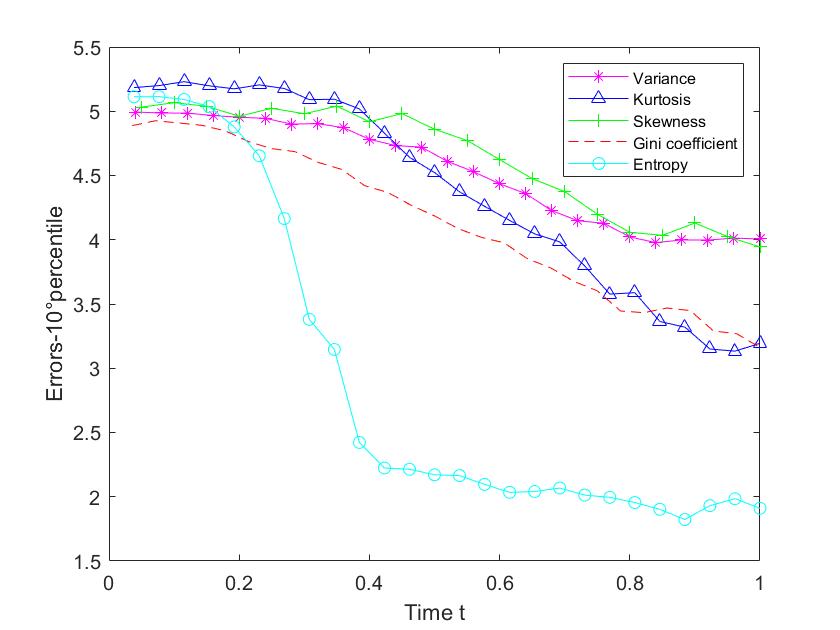}
{\small A)\par}
\end{minipage}
\ \hspace{5mm} \hspace{10mm} \
\begin{minipage}[b]{3cm}
\centering

\includegraphics[width=5cm]{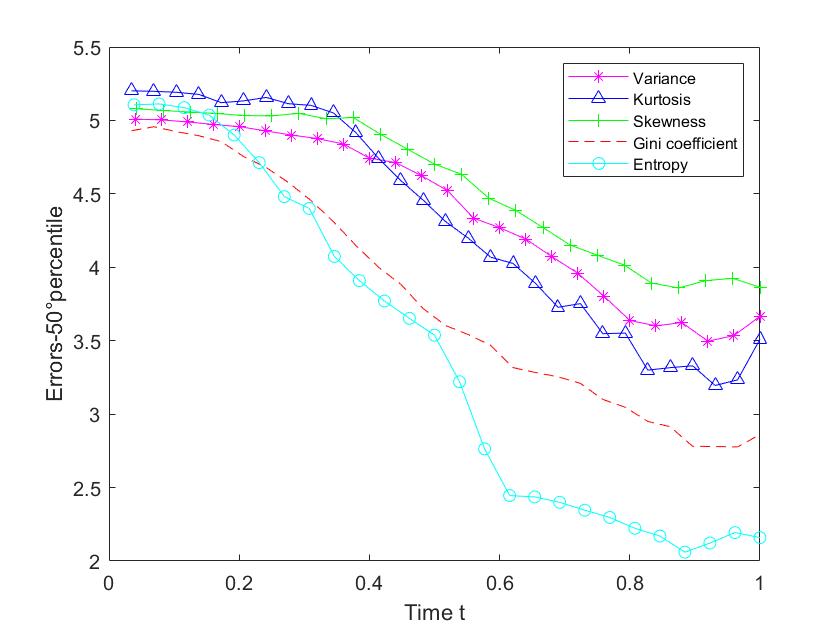}
{\small B)\par}
\end{minipage}
\ \hspace{5mm} \hspace{10mm} \
\begin{minipage}[b]{3cm}
\centering
\includegraphics[width=5cm]{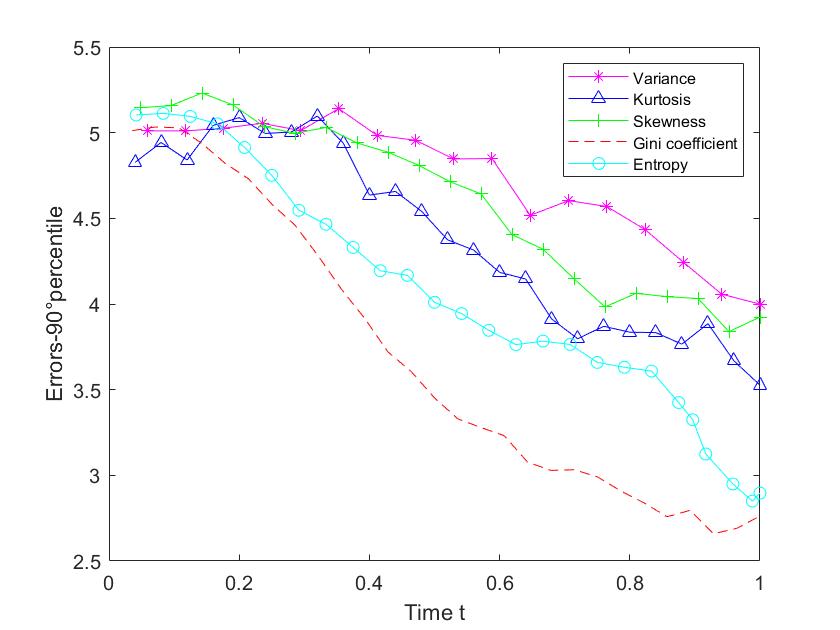}
{\small C)\par}
\end{minipage}

\captionstyle{normal}
\caption{Beta distribution of initial weights with $a=1$ and $b=3$. See caption to Fig. \ref{2} for further explanation}
\label{3}
\end{figure}

\begin{figure}
\begin{minipage}[b]{3cm}
\centering
\includegraphics[width=5cm]{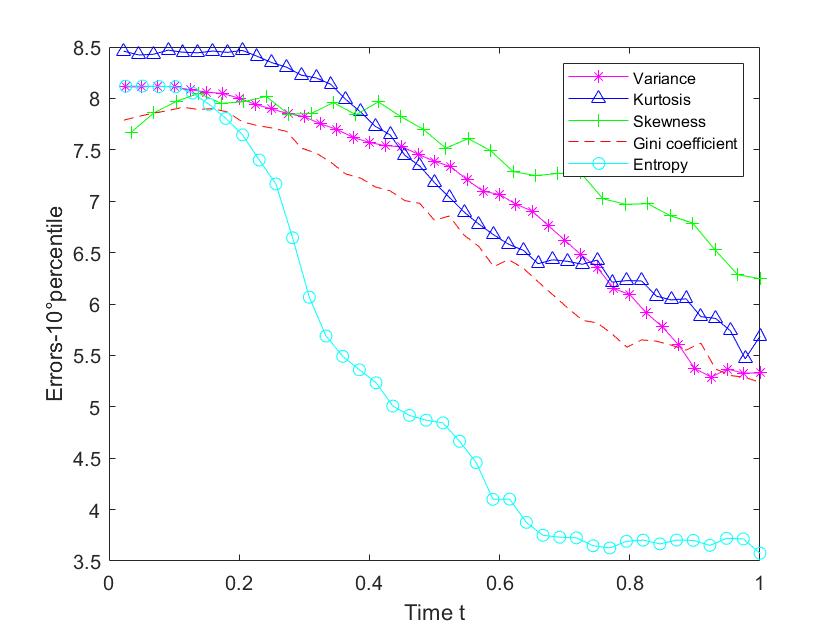}
{\small A)\par}
\end{minipage}
\ \hspace{5mm} \hspace{10mm} \
\begin{minipage}[b]{3cm}
\centering

\includegraphics[width=5cm]{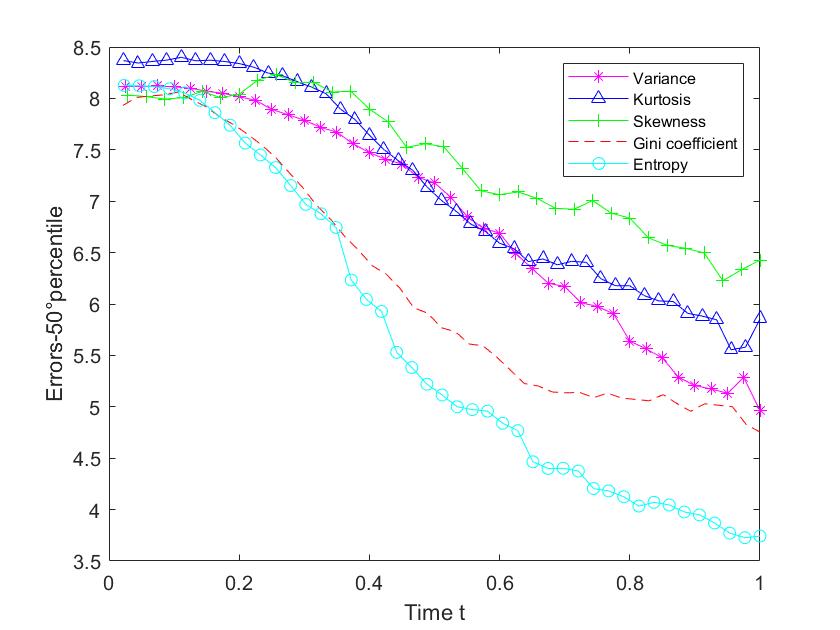}
{\small B)\par}
\end{minipage}
\ \hspace{5mm} \hspace{10mm} \
\begin{minipage}[b]{3cm}
\centering
\includegraphics[width=5cm]{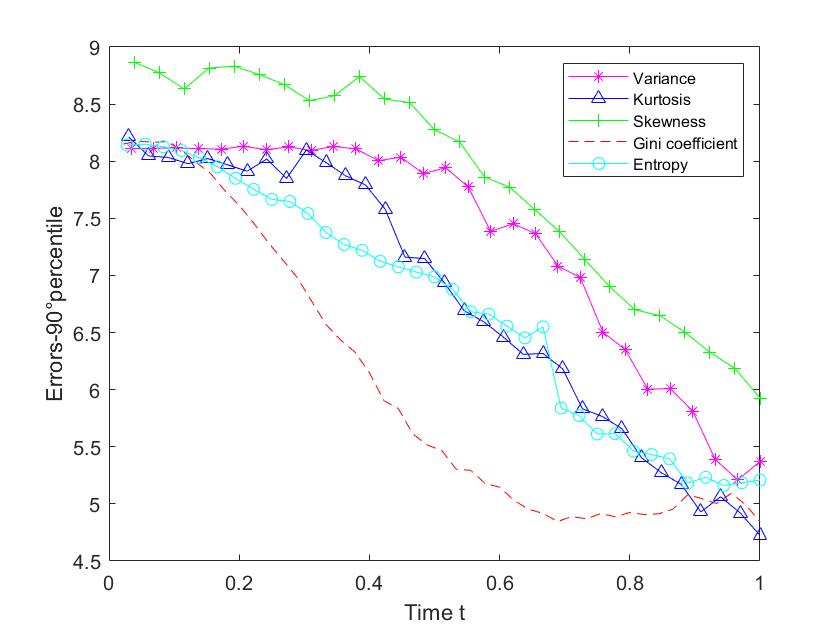}
{\small C)\par}
\end{minipage}

\captionstyle{normal}
\caption{Beta distribution of initial weights with $a=b=0.5$.
See caption to Fig. \ref{2} for further explanation}
\label{4}
\end{figure}

\begin{figure}
\begin{minipage}[b]{3cm}
\centering
\includegraphics[width=5cm]{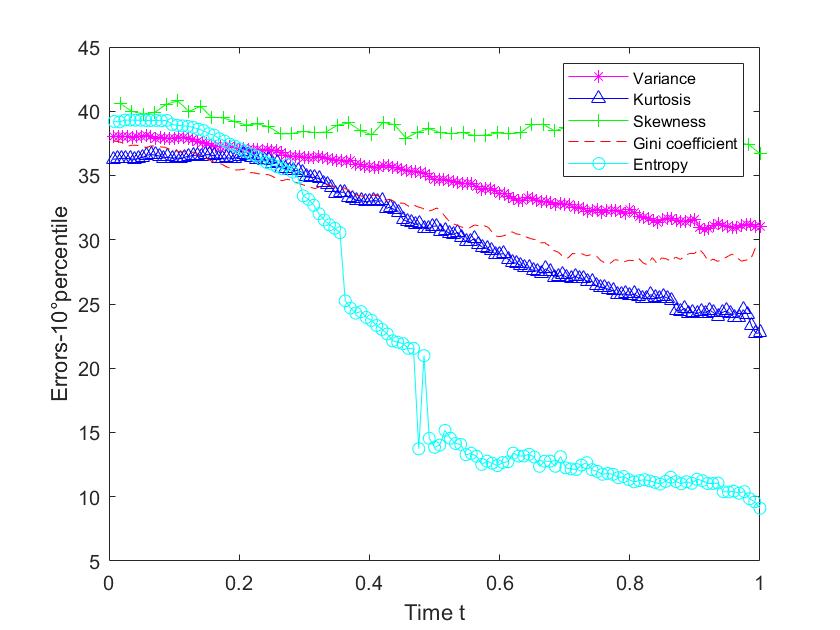}
{\small A)\par}
\end{minipage}
\ \hspace{5mm} \hspace{10mm} \
\begin{minipage}[b]{3cm}
\centering

\includegraphics[width=5cm]{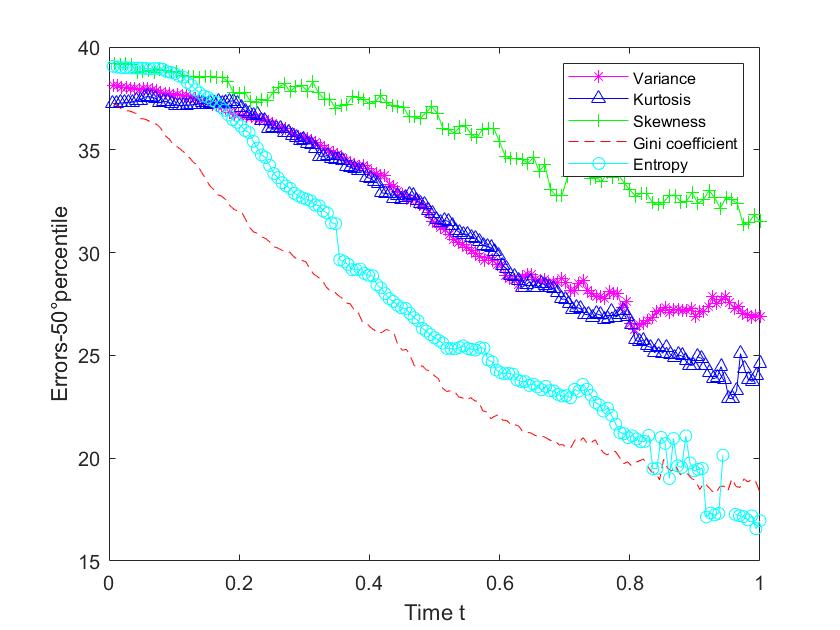}
{\small B)\par}
\end{minipage}
\ \hspace{5mm} \hspace{10mm} \
\begin{minipage}[b]{3cm}
\centering
\includegraphics[width=5cm]{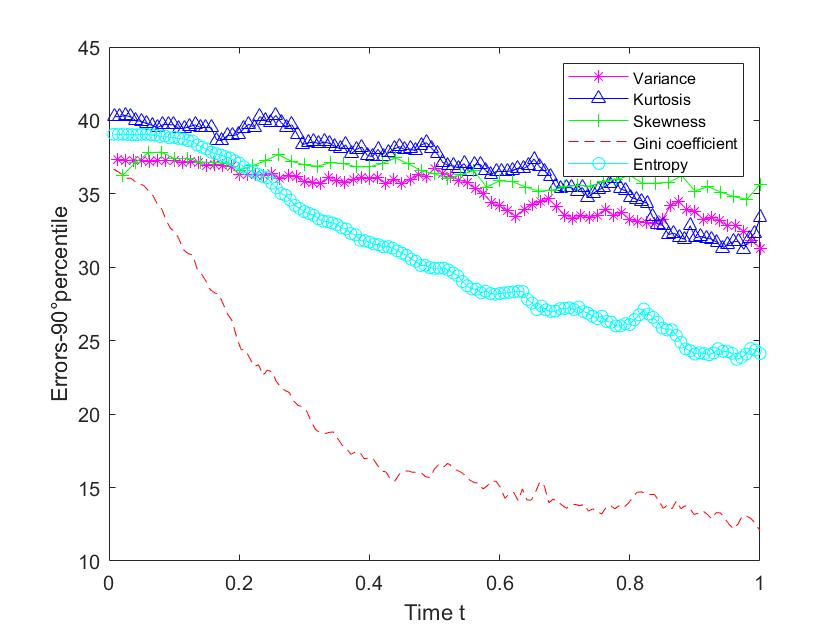}
{\small C)\par}
\end{minipage}

\captionstyle{normal}
\caption{Beta distribution of initial weights with $a=b=2$.
See caption to Fig. \ref{2} for further explanation}
\label{5}
\end{figure}

\begin{figure}
\begin{minipage}[b]{3cm}
\centering
\includegraphics[width=5cm]{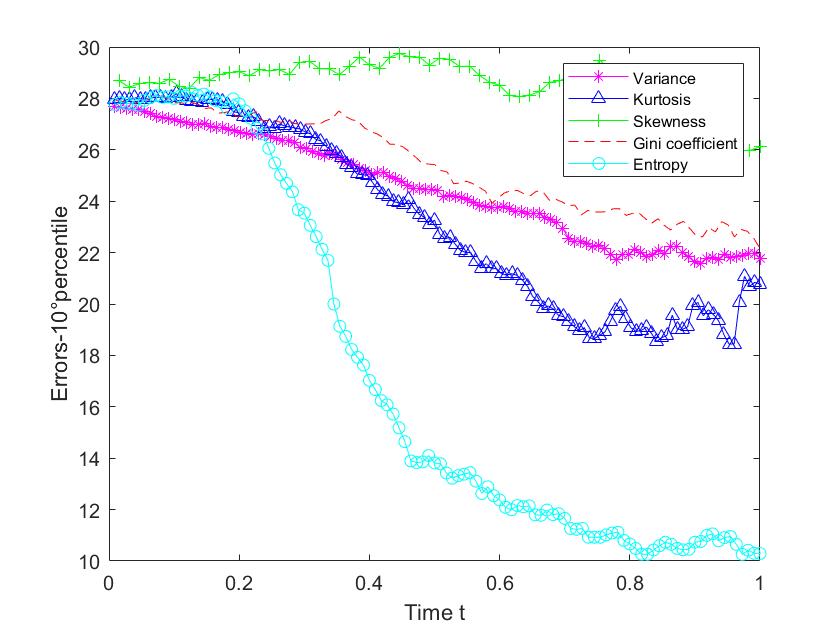}
{\small A)\par}
\end{minipage}
\ \hspace{5mm} \hspace{10mm} \
\begin{minipage}[b]{3cm}
\centering

\includegraphics[width=5cm]{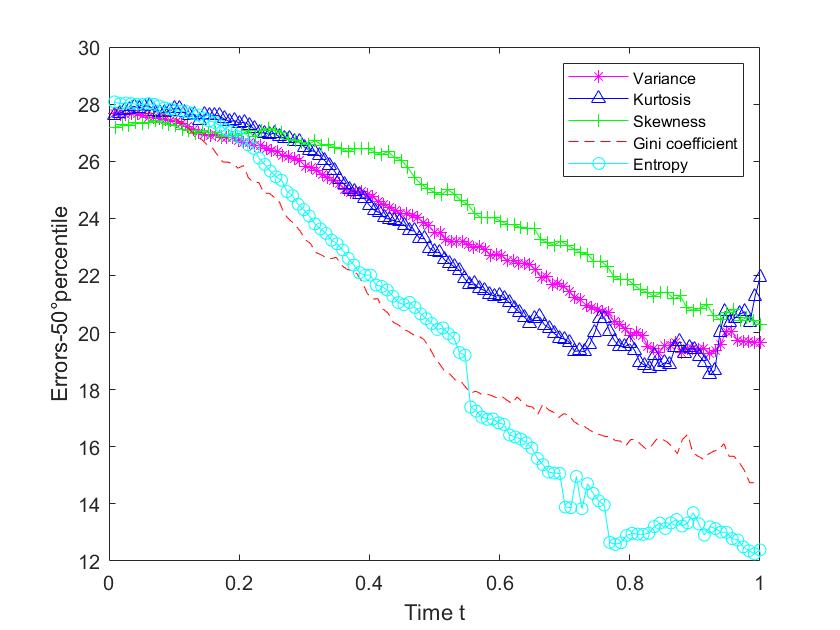}
{\small B)\par}
\end{minipage}
\ \hspace{5mm} \hspace{10mm} \
\begin{minipage}[b]{3cm}
\centering
\includegraphics[width=5cm]{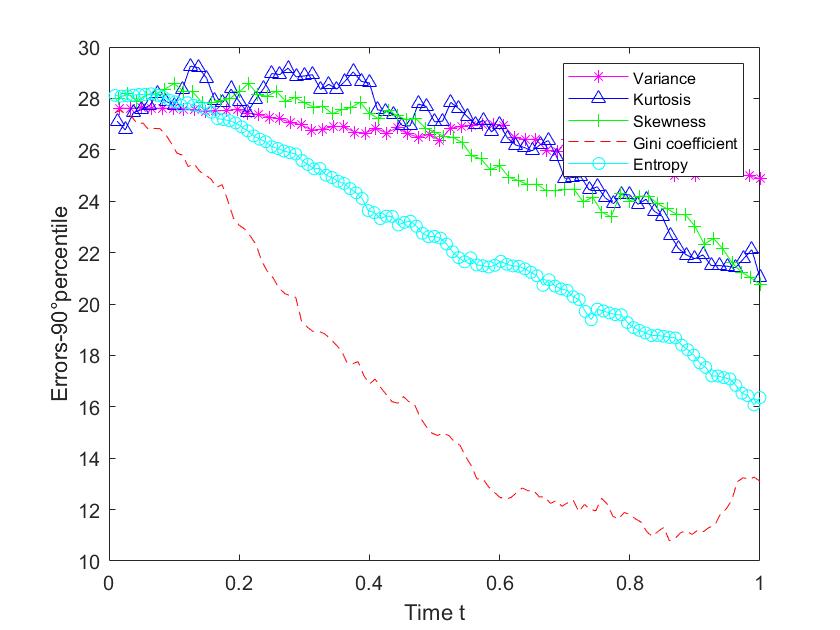}
{\small C)\par}
\end{minipage}

\captionstyle{normal}
\caption{Beta distribution of initial weights with $a=1$ and $b=0.5$. See caption to Fig. \ref{2} for further explanation}
\label{6}
\end{figure}

The main results of our method are all encompassed in Figures \ref{2}--\ref{6}. Each figure has three panels A--C, corresponding to the percentiles $per=10,50,90$. For each plot, the $y$-axis shows the error in the prediction given by (\ref{error}), while the $x$-axis shows the time. The five different graphs correspond to the five initial distributions of the weights. 


To allow an intuitive analysis of the predictive performance of the five statistical indicators $f$ and compare the systems, time is normalized to unity at failure as explained in the last section.

The straight line in each plot corresponds to the performance of the benchmark prediction \eqref{benchmark}.

A quick look at all three panels in all five figures gives an intuitive confirmation of how well our method works: in all five figures, in all three panels, and for all five functions our predictions are better than the benchmark at almost any time $t$. 
More specifically, the only clear but very short period where it would be better to use the benchmark prediction would be if the initial weights were randomly distributed, in the case of the 50\% percentile of the variance in the prediction of the in vivo systems. In this special case, we observe that the benchmark initially outperforms the rational expectations predictions; however, for longer times -- even in this extreme case -- the rational expectations predictions beat the benchmark predictions.

Moreover, the rational expectations predictions with {\it any\/} function $f$ become greatly superior to the benchmark prediction for times that are not too far away from failure. This outcome remains valid for any of the five considered distributions of the initial weights of the nodes. This is a clear illustration that the more information you have of a given in vivo system, the better your prediction can be. In contrast, close to the starting time -- e.g., if we consider the case where only one node has yet been broken -- all the in vivo systems look similar. In this case, the difference in performance of the various functions $f$ is negligible, whatever initial weight distribution is selected. In short, the best you can do without the benefit of any further information is to make the naive benchmark prediction. But as the in vivo system begins to deteriorate, every additional broken node provides new information, and that information should be used to optimize the prediction.

It is clear that, because the information obtained by each of the five possible functions $f$ will be different, it will lead to different possibilities for optimizing predictions. Consider the case of initially uniformly distributed weights shown in Fig. \ref{2}. Note that, for systems with abnormally small values of the statistical indicator (the 10\% percentile), the Shannon entropy gives predictions of the failure times which are around 80\% of the time better than those given by the other measures described by $f$. However, the particular in vivo system that is observed to deteriorate may not be one with a small Shannon entropy. It may instead have a persistently large Gini coefficient, say belonging to the 90\% percentile. In this case, it is the Gini coefficient that provides the optimal predictions.

This finding illustrates that as time goes by one should opportunistically switch between the different measures introduced by the five choices for $f$. This combined approach will ensure the path to globally optimal predictions, i.e., ones that are optimal at any given time $t$.

So what happens for the other initial weight distributions? Figure \ref{3} shows that the case of the Beta distribution with $a=1$ and $b=3$ has many features similar to the uniform distribution in Fig. \ref{2}, such as the Shannon entropy being the best function for making predictions for the 10\% and 50\% percentiles, while the Gini coefficient is the best for the 90\% percentile. So the fact of having a system with a higher initial concentration of small nodes, like the Beta distribution with $a=1$ and $b=3$, does not seem to give rise to a profoundly different optimal choice in our rational expectations predictions.

Consider then the more extreme case of systems that tend to have clusters of both small and large nodes, but few nodes of moderate size, corresponding to the Beta distribution with $a=b=0.5$, illustrated in Fig. \ref{4}. Again the Shannon entropy works well for the 10\% and 50\% percentiles, but now the 90\% percentile of the kurtosis predicts just as well as the 90\% percentile of the Shannon entropy, while it is still the 90\% percentile of the Gini coefficient that performs best.

For systems that have their nodes concentrated around their mean, corresponding to the Beta distribution with $a=b=2$ shown in Fig. \ref{5}, the Shannon entropy is only performs best for small percentiles, while the Gini coefficient is better for the 50\% and 90\% percentiles. Somewhat surprisingly, for systems that have their nodes concentrated around large values, corresponding to the Beta distribution with $a=1$ and $b=0.5$ shown in Fig. \ref{6}, performance does not differ notably from the uniform distribution (Fig. \ref{2}).

We note that the Gini coefficient is not usually used as a performance measure in failure prediction models. One exception is \cite{Ooghe}, whose authors would agree with us that the Gini coefficient is a powerful and attractive measure.
In the literature, several studies have been carried out in the field of reliability theory -- specifically, considering $k$-out-of-$n$ systems -- which exploit the various statistical moments to validate their models. We refer for example to Reijns and Gemund \cite {Reijns} or Amari et al. \cite {Amari}. According to these authors, lower moments are more robust than higher moments. In contrast, our results seem in general to indicate an advantage in using the kurthosis rather than the variance, and both the variance and the kurthosis seem in general to perform better than the skewness. However, there are exceptions, notably for the asymmetric distributions which allow large values -- i.e., the Beta distribution with $a=1$ $b=0.5$. For the 90\% percentile for these two distributions, the skewness equals or outperforms both the variance and the kurtosis. This makes sense since knowing for an asymmetric distribution that the remaining set of nodes have a large skewness indicates a system that has a large fraction of nodes with large values. This information is useful in predicting the (probably longer than usual) failure time.

\section{Conclusions}
\label{sec6}

In this paper, we have shown that using rational expectations -- together with the idea of using different measures that can change over time -- enables us to come up with tools for failure predictions for weighted $k$-out-of-$n$ systems. The core idea in our method is that, every time a node breaks, this provides new information, and one should use this information via rational expectations to continually optimize predictions.

Through extensive computer simulations, we have shown how our new method can dynamically outperform static predictions. We have explored how the procedure works for systems with five different initial distributions of the weights of the nodes. To see how different information influences the predictions for systems with different initial distributions of the weights, we have tested five measures: mean, variance, kurtosis, Gini coefficient, and Shannon entropy. As we have shown, different measures may be optimal depending on the given initial distribution of a system. However, as we have also shown, we can obtain optimal predictions by adopting a dynamic perspective -- viz., by switching between different measures at different times, depending on the state of deterioration of the given system.

The presented framework is general and can be suitably modified and adapted to several contexts. In particular, it is easy to modify the method to apply it for any given initial distribution of weights. Obviously, the five measures we introduced were meant only to illustrate how using different information might be optimal depending on the state of the system and the initial distribution of the weights. Our claim is not that these measures are ``the best'', but rather to illustrate their different impact on the accuracy of prediction. Better measures might possibly be found for different circumstances.

We argue that our method could have real practical relevance in economics and finance, for example, for banking networks or for assessing the systemic risk of a country, the Eurozone, or sovereign credit, among other things. Our future research will focus on such practical examples.

Actually, the rational expectations prediction method should lead to a broad range of studies. A natural extension would be to try out other reallocation rules instead of the proportional one; for other systems, it may be more natural to apply a uniform reallocation rule or a threshold-based reallocation rule. One could also consider rules that depend on the trajectories of previous failures.
And as mentioned beforehand, the analysis could be extended by comparing additional statistical indicators: the Frosini index to compare with the Gini coefficient, the Bienaym--Chebyshev inequality, the Pearson index to capture the dependence between the components, and Goodman and Kruskal's index, taking into account the correlation between the components. Other measures belonging to the Shannon family of entropies could also be studied, such as the Kullback--Leibler divergence, the Jeffreys distance, the K divergence, or the Jensen difference. In short, we encourage the reader to use our method to study the performance of other measures for their specific system of choice.


\end{document}